\documentstyle[12pt]{article}
\newcommand{\doublespace}{
   \renewcommand{\baselinestretch}{1.5}
   \large\normalsize}

\renewcommand{\u}[1]{${\bf #1}$}
\newcommand{\be}{\begin{equation}}
\newcommand{\by}{\begin{eqnarray}}
\newcommand{\ee}{\end{equation}}
\newcommand{\ey}{\end{eqnarray}}

\newcommand{\Sp}{\makebox[.5in]{}}
\newcommand{\ra}{\rightarrow}

\renewcommand{\thesection}{\arabic{section}}

\begin{document}
\doublespace

\begin{center}
\large {\bf  Optimizing the RVB state on a  triangular  lattice:\\
Presence of the long-range order}\\
\normalsize Yong-Cong  Chen\\
Department of Mathematics,  Rutgers University\\ New Brunswick, NJ 08903\\
and\\
$^{\dagger}$Department of Physics\\University of Science
\& Technology of China\\
Hefei, Anhui 230026, China\\
\vspace{.15in}
\end{center}

\newcommand{\bff}[1]{{\bf #1}}
\newcommand{\bk}{{\bff k}}
\begin{center} {\bff Abstract}\end{center}
We present a Schwinger-boson approach  for the RVB state of the
spin-1/2  Heisenberg antiferromagnet on a triangular lattice. It is
shown that a Gutzwiller  projection of the mean-field state that
includes {\em both}  antiferromagnetic and ferromagnetic decouplings
leads to optimizing  the RVB pair amplitudes within a self-consistent
approximation. The resulting state yields, by  Monte Carlo
simulations, energies and spin-spin correlations in
excellent agreement with  the exact diagonalization result
on finite lattices (up to 36 sites). We conclude  that the
{\em optimized RVB wavefunction} possesses a long-range
three-sublattice order.

\vspace{.1in}

\noindent PACS numbers: 75.10.Jm, 75.30.Ds, 74.70.Vy




\vspace{1.5in}

\noindent $^{\dagger}$Permanent address.

\newpage
\renewcommand{\d}{\dagger}
\newcommand{\ua}{\uparrow}
\newcommand{\da}{\downarrow}
\newcommand{\Tr}{\mbox{Tr}}


The spin-1/2 antiferromagnetic Heisenberg model on a triangular lattice
has been a subject of great interest in recent years.
Unlike the square lattice case,
the corresponding Ising model is disordered at all temperature with
finite entropy\cite{class}. This was also the original model for
Anderson's  proposed resonant valence bond (RVB) state\cite{PWA}.
The system  was thereafter often refered to be in  quantum  liquid
state  for the lack of long-range magnetic order (LRMO).
However, this simple picture was not so obvious: The quantum
fluctuations may as well lift the classical degeneracies and lead to ordered
ground-state. Many studies were further inspired  by the
magnetic  properties of doped $CuO_{2}$ planes in the copper-oxide
superconductors where the frustration again plays a central role.
Almost every possible scheme has been employed in the literature.
These include the earlier variational
approaches\cite{PWA,KL,Huse},  the spinwave and large-$N$
theories\cite{spin,Large-N}, the spontaneous symmetry breaking
consideration\cite{SSB}, the numerical series
expansion methods\cite{Ising,high-T}, and the exact
diagonalizations on finite clusters\cite{exact0,exact1,exact2}.
Most of recent analytic aproaches  tend to conclude with
LRMO. But they all seem to suffer from  uncontrolled
approximations or inaccuracies. On the other hand, the numerical groups
are  divided into with and without LRMO (or just right on the critical
point as suggested by Singh and Huse\cite{Ising}), mainly due to insufficient
knowledge on  larger clusters and different methods of extrapolation.
 The main question,  whether  this frustrated system possesses LRMO
remains controversial.

Even  RVB states could have LMRO if the bonds decayed sufficiently
slowly. This was first found on a square lattice  by Liang, Doucot
and  Anderson\cite{LDA}. Does this also take place  on the the
triangular  lattice? In this Letter, we shall address to this question
based on a new use of the Schwinger-boson formalism\cite{chen1,chen2}.
The central idea is to work with the RVB
states resulting from  Gutzwiller projection of mean-field states onto
the right Hilbert  space. This method has been working extremely well on the
square lattice\cite{chen2}.  Note only the  ground-state energy
and the staggered magnetization obtained are virtually identical to the exact
ones (cited by\cite{review}), the low lying spin-flip spectrum
also agrees with the best numerical result found by
supercomputer (cf. \cite{Super-C}).  On the triangular lattice, one needs to
face the frustration.   we shall show that it is important  to have,
at the mean-field level, both the antiferromagnetic and
ferromagnetic  fields in order to account for this frustration.
With a proper balance between the two, the Gutzwiller projection then
leads to optimizing the RVB state.  The latter is also explicitly derived
within a self-consistent approximation in the ``loop-gas''
statistics\cite{Suth,LDA}. Our approach provides a simple but powerful
way of founding the best RVB state and  the right
mean-field solution {\em regardless of  the frustration}. The
optimized parameter-free RVB state is  found to have energies much
closer to the  exact ones than those previously found\cite{KL,Huse}.
The system  is predicted to be long-range ordered with a staggered
sublattice magnetization $\cong 0.275$.


The nearest neighbor antiferromagnetic Heisenberg model  may be written in
terms of Schwinger bosons as\cite{AA}
\be\label{e1}
\hat{H}=\sum_{<ij>}\left[-\frac{(1+\nu)}{4}
\hat{A}_{ij}^{\d}\hat{A}_{ij}+
\frac{(1-\nu)}{4}\hat{F}^{\d}_{ij}\hat{F}_{ij}+\nu S^{2}
-\frac{(1-\nu)}{2}S\right],
\ee
where
\be\label{e2}
\hat{A}_{ij}=\hat{b}_{i\ua}\hat{b}_{j\da}-\hat{b}_{i\da}\hat{b}_{j\ua},\Sp
\hat{F}_{ij}=\sum_{\sigma=\ua\da}\hat{b}^{\d}_{i\sigma}\hat{b}_{j\sigma},\Sp
\sum_{\sigma=\ua\da}\hat{b}^{\d}_{i\sigma}\hat{b}_{i\sigma}=2S.
\ee
We have deliberately broken up in (\ref{e1}) the Hamiltonian into  two
pieces (the first two terms): The first (second) is  suitable  for
an antiferromagnetic (ferromagnetic) mean-field decoupling.  The
separation  ($\nu$ is a free parameter to be  fixed)  is  motivated by
the fact that the pairing of  nearest neighbor  spins is frustrated on
the triangle lattice. A ferromagnetic field will therefore be
essential for an accurate description of the original system.  This
will become clear below when we
minimize the energy of the RVB state.
The partition function of the system can be calculated via
\[
Z(\beta)=\Tr[\hat{{\cal P}}_{G}\hat{\rho}]
=\Tr\left\{ [\prod_{i=1}^{N}\hat{P}_{i}] \hat{\rho}\right\},
\]
where $\hat{\rho}=\exp(-\beta\hat{H})$, and
$\hat{{\cal P}}_{G}$ ($\hat{P}_{i}$)  stands for the Gutzwiller
projection operator for the whole lattice (the $i$th site)
which enforces the constraint in (\ref{e2}).
Introduce the so-called coherent state, $|\alpha_{i}>=$
 $\exp\left(\sum_{\sigma=\ua\da}\hat{b}^{\d}_{i\sigma}
b_{i\sigma}\right)|0>$, where
$\alpha_{i}=(b_{i\ua},b_{i\da})$ are ordinary complex
numbers. We can further write\cite{chen1}
\be\label{e4}
Z(\beta)=\left.\left[\prod_{i=1}^{N}P_{i}\right]
<\{{\alpha}_{i}^{\d}\}|\hat{\rho}|\{\alpha_{i}\}>
\right|_{\{{\alpha}^{\d}_{i},\alpha_{i}=0\}},
\ee
\be\label{e5}
P_{i}=\frac{1}{(2S)!}\left[\sum_{\sigma=\ua\da}
\frac{\partial}{\partial b_{i\sigma}}
\frac{\partial}{\partial {b}^{\d}_{i\sigma}}\right]^{2S}
\ee
provided that the matrix elements of $\hat{\rho}$ are known.

Eq. (\ref{e4}) would be rigorous if the exact matrix elements were
used. This is of course rarely possible for interacting systems.
One therefore needs a trial $\hat{\rho}$. We shall consider in this
Letter a mean-field type of approximation in which the Hamiltonian is
replaced by
\[
\hat{H}_{\mbox{mf}}=E_{0}+
\lambda\sum_{i,\sigma}\hat{b}^{\d}_{i\sigma}\hat{b}_{i\sigma}+
\sum_{<ij>}\left\{[-D_{ij}\hat{A}^{\d}_{ij}+
Q_{ij}\hat{F}_{ij}]+h.c.\right\}.
\]
The Lagrangian multiplier $\lambda$ and the  mean fields $D_{ij}$ and
$F_{ij}$ are to be determined via the usual
self-consistency requirements,
\be\label{e7}
\sum_{\sigma=\ua\da}<\hat{b}_{i\sigma}^{\d}
\hat{b}_{i\sigma}>=1,  \Sp
D_{ij}=\frac{1+\nu}{4}<\hat{A}_{ij}>, \Sp
Q_{ij}=\frac{1-\nu}{4}<\hat{F}^{\d}_{ij}>.
\ee
$\hat{H}_{\mbox{mf}}$ after Fourier
transforms can be readily diagonalized by a Bogoliubov transformation.
\newcommand{\br}{{\bff r}}
This  results in a  mean-field spectrum $\omega_{\bk}+M_{\bk}$  with
\[
\omega_{\bk}=\sqrt{\lambda_{\bk}^{2}-|D_{\bk}|^{2}},\Sp
M_{\bk}=(Q_{-\bk}-Q_{\bk})/2,\Sp
\lambda_{\bk}=\lambda+(Q_{-\bk}+Q_{\bk})/2.
\]
The matrix elements of $\hat{\rho}$ can be calculated accordingly.
We find\cite{chen1}
\[
<\{\alpha_{i}^{\d}\}|\hat{\rho}_{\mbox{mf}}|\{\alpha_{i}\}>
=Z_{0}(\beta)\exp\left(\sum_{\bk}\left[
\sum_{\sigma=\ua\da}W^{(1)}_{\bk}b^{\d}_{\bk\sigma}b_{\bk\sigma}
+(W^{(2)}_{\bk}
b^{\d}_{\bk\ua}b^{\d}_{-\bk\da}+c.c.)\right]\right)
\]
where
\[W^{(1)}_{\bk}=B_{\bk}\exp(-\beta M_{\bk}),\Sp
W^{(2)}_{\bk}=(D_{\bk}/\omega_{\bk})B_{\bk}\sinh(\beta
{\omega}_{\bk}),\]
\be\label{e12}
B_{\bk}=[(\lambda_{\bk}/\omega_{\bk})\sinh(\beta{\omega}_{\bk})
+\cosh(\beta{\omega}_{\bk})]^{-1}.
\ee
Note that $Q_{\bk}=Q_{\bk}^{\ast}$, $D_{\bk}=-D_{-\bk}$.


Now return to (\ref{e4}) and consider in what follows
$S=1/2$. Write $Z(\beta)=Z_{0}(\beta)\times Y_{N}$ and
take a closer look at the structure of
$Y_{N}$. The differentiations amount to dividing  the whole
lattice into subsets of self-avoiding loops\cite{chen1,chen2}.
One then sums over all possible configurations of the loops.
This kind of {\em loop gas}  statistics has been known\cite{LDA,Suth}
for the RVB states.  The more general case\cite{chen1,chen2} is that
the bond, say, from sites $i$ to $j$ is replaced by a
$4\times 4$ transfer matrix ${\bff T}_{ij}$ of the form
($\sigma_{z}$ below is the third Pauli matrix)
\be\label{e13}
{\bff T}_{ij}=\left(\begin{array}{cc} {\bff {G}}_{ij}
 & 0 \\ 0 & \sigma_{z}{\bff {G}}_{ij}\sigma_{z}\end{array}\right),
\Sp
{\bff {G}}_{ij}=\left(\begin{array}{cc}W^{(1)}_{ij} & -W^{(2)\ast}_{ij}
\\ W^{(2)}_{ij} &W^{(1)\ast}_{ij}\end{array}\right) .
\ee
It brings, upon taking $P_{i}$, see (\ref{e5}),
the prefactor $(\chi_{i1}b_{i\ua}+\chi_{i2}b^{\d}_{i\da}+
\chi_{i3}b_{i\da}+\chi_{i4}b_{i\ua}^{\d})$ to a new one at site $j$
via multiplying ${\bff T}_{ij}$ to the column of the $\chi$'s.
$W_{ij}^{(1)}$, $W_{ij}^{(2)}$
are, respectively, the (inverse) Fourier transforms of $W_{\bk}^{(1)}$,
$W_{\bk}^{(2)}$. Tracing over the matrix (divided by two) after completing
a loop gives its contribution.

To illustrate this prescription,
one can decompose $Y_{N}$ at a given $j_{0}$ into
(all $j_{k}$'s below are self-avoiding)
\be\label{e14}
Y_{N}=\sum_{n=0}^{N-1}\left\{\sum_{\{j_{k}; k\neq 0\}}
Y_{N-n-1}(\{j_{k}\})\times
\left[\mbox{Tr}({\bff {G}}_{j_{0}j_{1}}\cdots
{\bff {G}}_{j_{n}j_{0}})\right]\right\}.
\ee
The arguments of $Y_{N-n-1}$ represent the sites excluded.
Various physical quantities can be calculated by
multiplying the operators to $\hat{\rho}$ and then compute the matrix
elements as in (\ref{e4}).
In the coherent states one has the simple substitutions,
$\hat{b}^{\d}_{i\sigma}\ra {b}^{\d}_{i\sigma}$,
$\hat{b}_{i\sigma}\ra (\partial /\partial b^{\d}_{i\sigma})$,
 which may modify the transfer matrix at site $i$.
In general, operators that conserve boson numbers
can be presented by some $4\times 4$ matrices inserted at the relevant
sites.  One then can proceed to evaluate the modified $Y_{N}$ as usual.
For the spin-spin correlation
$\hat{{\bff S}}_{i}\cdot\hat{{\bff S}}_{j}$ we have the  matrices
\[\frac{1}{4}
\left(\begin{array}{cc}\sigma_{z}& 0\\ 0 & -\sigma_{z}\end{array}\right)_{i}
\left(\begin{array}{cc} \sigma_{z} & 0\\ 0 & -\sigma_{z}\end{array}\right)_{j}
+\frac{1}{2}\left[
\left(\begin{array}{cc}0 & 0\\ {\bff I} &0 \end{array}\right)_{i}
\left(\begin{array}{cc} 0& {\bff I}\\  0 & 0 \end{array}\right)_{j}
+(i\leftrightarrow j)
\right].\]
This leads us to ($i\neq j$)
\be\label{e15}
\left.<\hat{{\bff S}}_{i}\cdot\hat{{\bff S}}_{j}>
=\frac{1}{Y_{N}}\sum_{n=1}^{N-1}\left\{
\sum_{\{j_{k}\}}
Y_{N-n-1}(\{j_{k}\})\times\frac{3}{4}\sum_{m=1}^{n}
\mbox{Tr}[\sigma_{z}{{\bff G}}^{(1)}_{j_{0}j_{m}}
\sigma_{z}{{\bff G}}^{(2)}_{j_{m}j_{0}}]
\right\}\right|_{j_{0}=i,j_{m}=j},
\ee
\be\label{e16}
{{\bff G}}^{(1)}_{j_{0}j_{m}}\equiv {\bff G}_{j_{0}j_{1}}\cdots
{\bff G}_{j_{m-1}j_{m}}, \Sp
{{\bff G}}^{(2)}_{j_{m}j_{0}}\equiv {\bff G}_{j_{m}j_{m+1}}\cdots
{\bff G}_{j_{n}j_{0}}.
\ee


We shall take here  a specific  mean-field solution
Consider the simplest symmetry  $Q_{ij}=Q$,
$D_{ij}=iD\exp(i3\phi_{ij})$ with $\phi_{ij}$ being the polar angle of
the bond.  Let
$\gamma_{\bk}=(1/3)\sum_{i=1}^{3}\cos(\bk\cdot{\bff e}_{i})$,
$\Gamma_{\bk}=(1/3)\sum_{i=1}^{3}\sin(\bk\cdot{\bff e}_{i})$
where
${\bff e}_{1}=(1/2,-\sqrt{3}/2)$,
${\bff e}_{2}=(1/2,+\sqrt{3}/2)$, and
${\bff e}_{3}=(-1,0)$.
At zero temperature, the mean-field dispersion
$\omega_{\bk}/\lambda=\tilde{\omega}_{\bk}$ =
$\sqrt{(1+d_{1}\gamma_{\bk})^{2}-(d_{2}\Gamma_{\bk})^{2}}$ becomes gapless.
Namely we have ($z=6$ is the coordination number)
\be\label{e21}
1+d_{1}\gamma_{\pm\bk_{0}}=\pm d_{2}\Gamma_{\pm\bk_{0}},
\Sp  d_{1}=zQ/\lambda, \Sp d_{2}=zD/\lambda.
\ee
where $\bk_{0}=2\pi/3(1,\sqrt{3})$.  Condensation, denoted by
$\alpha$ which has to be inserted in (\ref{e7}), 
occurs at $\pm \bk_{0}$ [or their equivalences on the
reciprocal lattice spanned by   ${\bff G}_{1}=\pi(2,\sqrt{3})$ and ${\bf
G}_{2}=(0,4\pi/\sqrt{3})$].
Consequently, the mean-field equations of (\ref{e7}) read (``BZ'' means the
first Brillouin zone)
\[
1=\alpha+\frac{1}{2}\int_{BZ}\frac{\sqrt{3}d^{2}k}{8\pi^{2}}
\left[\frac{1+d_{1}\gamma_{\bk}}{\tilde{\omega}_{\bk}}\right],\Sp
\]
\[
\frac{D}{1+\nu}=\frac{\alpha}{2}\Gamma_{\bk_{0}}+\frac{1}{4}
\int_{BZ}\frac{\sqrt{3}d^{2}k}{8\pi^{2}}\left[\frac{d_{2}\Gamma_{\bk}^{2}}{
\tilde{\omega}_{\bk}}\right],
\]
\be\label{e24}
\Sp\frac{Q}{1-\nu}=\frac{\alpha}{2}\gamma_{\bk_{0}}+\frac{1}{4}
\int_{BZ}\frac{\sqrt{3}d^{2}k}{8\pi^{2}}\left[
\frac{\gamma_{\bk}(1+d_{1}\gamma_{\bk})}{
\tilde{\omega}_{\bk}}\right].
\ee
The matrix elements $W_{ij}$'s  can be found upon
substituting the the solution  back to (\ref{e12}), see (\ref{e31}) below.


We next show that this set of $W_{ij}$'s minimizes  the RVB ground-state
energy at $\nu=0$ within a self-consistent approximation for the
loop gas. Going back  to the transfer matrix (\ref{e13}),
$W_{\bk}^{(1)}$ vanishes at zero temperature.  The physical picture of
(\ref{e13})--(\ref{e16}) become apparent: Here $W^{(2)}_{ij}$ stands for
the amplitude of a RVB pair connecting site $i$ to  $j$, more explicitly,
$W_{ij}(\hat{b}^{\d}_{i\ua}\hat{d}^{\d}_{j\da}-
\hat{b}^{\d}_{i\da}\hat{d}^{\d}_{j\ua}$
[the superscript ``(2)'' is hereafter
dropped].  $Y_{N}$ is nothing but $<\Phi_{\mbox{RVB}}|
\Phi_{\mbox{RVB}}>$ with $|\Phi_{\mbox{RVB}}>$ containing all possible
superpositions of the pair products. Eq. (\ref{e15}) then gives  the
explicit way of computing the spin-spin correlations.  The
question is reduced to finding the set of best amplitudes that
minimizes the energy.

Rigorous evaluation of the RVB expectation values appears problematic.
We thus introduce below a self-consistency approach.  Let us first
approximate $Y_{N}/Y_{N-n-1}\rightarrow y^{n+1}$.  This assigns a
uniform weight $1/y^{n+1}$ for a loop of $(n+1)$ sites. We are still
left to deal with the self-avoiding restriction.  Let us  try to ignore
this in the first place, denoting the corresponding $y$ by $y_{0}$.
The correct $y$ is then obtained by properly identifying the
over-counting.  This allows a full analytic summation over the loops
using Fourier transforms. The  matrix  after summing over the
paths  connecting $i$ to $j$  reads [$i\neq j$, see (\ref{e16})],
\be\label{e25}
{\bff R}_{ij}
=\frac{1}{N}\sum_{\bk}\frac{\exp(-i\bk\cdot\br_{ij})
}{1-|W_{\bk}/y_{0}|^{2}}
\left(\begin{array}{cc} 1 &W^{\ast}_{\bk}/y_{0}\\ W_{\bk}/y_{0} &1
\end{array}\right).
\ee
The parameter $y_{0}$ can be determined by Tr$({\bff R}_{ii}-1)=1$ [cf.
(\ref{e30}) below]. Now  $y$ can be approximately recovered by
grouping the extra winding  at a given site (say $i$)  into $1/y$,
which gives
\be\label{e26}
y^{-1}=y_{0}^{-1}\times\mbox{Tr}{\bff R}_{ii}/2
=(3/2)\times y_{0}^{-1}.
\ee
More detail on this over-counting problem has been dicussed in \cite{chen2}.

The same procedure can be applied to the spin-spin correlations.
We now have one path starting from $i$ to $j$ and the other from
$j$ to  $i$.  Ingoring again the self-avoiding restriction and the overlaps
between the two paths, the result of the summation is  simply
${\bff R}_{ij}$ and ${\bff R}_{ji}$.
However, taking  two independent paths results in  an additional
over-counting of a factor 3/2 [see
(\ref{e26})]  which has to  be deducted.  Sutbstituting (\ref{e25})
back to (\ref{e15}) and
assuming, as in the mean-field case, the three-fold
symmetry for $W_{\bk}$, we end up with the expression (drop $y_{0}$
hereafter as $W_{\bk}$ is variational)
\[
E_{\mbox{bond}}=\left|\frac{1}{N}\sum_{\bk}\frac{
\gamma_{\bk}}{1-|W_{\bk}|^{2}}\right|^{2}
-\left|\frac{1}{N}\sum_{\bk}\frac{\Gamma_{\bk}W_{\bk}}{
1-|W_{\bk}|^{2}}\right|^{2}
\]
subject to $W_{\bk}=W_{\bk}^{\ast}=-W_{-\bk}$ and the constraint
\be\label{e30}
\frac{1}{N}\sum_{\bk}\frac{1}{1-|W_{\bk}|^{2}}=\frac{3}{2}.
\ee
Eq. (\ref{e30}) can be treated via the method of Lagrangian multiplier.
For simplicity, one  can assume here  real $W_{\bk}$'s.
It is now a straightforward exercise to obtain (picking up the right
solution satisfying $|W_{\bk}|\leq 1$)
\be\label{e31}
W_{\bk}=\frac{d_{2}\Gamma_{\bk}}{1+d_{1}\gamma_{\bk}+
\sqrt{(1+d_{1}\gamma_{\bk})^{2}-(d_{2}\Gamma_{\bk})^{2}}}
\ee
with
\be\label{e32}
\frac{d_{1}}{d_{2}}=\frac{(1/N)\sum_{\bk}[\gamma_{\bk}/(1-W_{\bk}^{2})]}{
(1/N)\sum_{\bk}[\Gamma_{\bk}W_{\bk}/(1-W_{\bk}^{2})]}.
\ee
Note that (\ref{e31}) is nothing but the zero-temperature limit of
(\ref{e12}) when the mean-field solution discussed above is
used. Moreover, it can be easily checked that  (\ref{e30}) and
(\ref{e32}),  after some manipulations using the specific form of
(\ref{e31}),  reduce to (\ref{e21}) and (\ref{e24}) of
the mean-field result with $\nu=0$. Note that, in passing to the
continous limit,  the same condensation
$\alpha$  is also required in (\ref{e30}) and (\ref{e32}).
This proves our claim that two approaches give the same RVB state.

Solving (\ref{e30}) and (\ref{e32}), or equivalently
(\ref{e21}) and (\ref{e24}),  we find
$\alpha=0.275$, $d_{1}=-0.7223$, $d_{2}=1.572$ and $E_{\mbox{bond}}=-0.1899$.
Because of the condensation at the particular modes $\pm\bk_{0}$, the system
possesses  a long-range three-sublattice order with
\be\label{e33}
\lim_{|\br_{ij}|\rightarrow\infty}
<\hat{{\bff S}}_{i}\cdot\hat{{\bff S}}_{j}>
=\left\{\begin{array}{cc}\alpha^{2} & i,j\in \mbox{same sublattices}\\
-\alpha^{2}/2 &i,j \in\mbox{different sublattices}\end{array} \right..
\ee


The structure of (\ref{e14}) and (\ref{e15}) allows also a direct
evaluation via Monte Carlo simulation.
{\em This is important for justifying  our theory.}
 It amounts to sampling over
the distribution of the loops.  At zero temperature, the rule to
calculate the spin-spin correlations is particularly simple:
For a given configuration,
\[
<\psi_{L}|\hat{{\bff S}}_{i}\cdot\hat{{\bff S}}_{j}|\psi_{R}>=
\left\{\begin{array}{ccc}-3/4 <\psi_{L}|\psi_{R}>
& i,j\in \mbox{same loop}&
n(i)-n(j)=\mbox{odd}\\
+3/4 <\psi_{L}|\psi_{R}>
& i,j\in \mbox{same loop}&
n(i)-n(j)=\mbox{even}\\
0 & i,j\in \mbox{different loops}&
\end{array} \right.
\]
where $n(k)$ stands for the order of the site $k$ in its own loop.
Each loop contains even number of sites.
In the simulation, the loop configurations are updated
by randomly choosing a pair of nearest or next nearest neighbor sites
and exchanging their loop connections with a probability satisfying
detailed balance condition\cite{LDA,chen2}.   For the triangular
lattice, the sign problem shows up in that some of the loops contribute
negative weights. We use their absolute values
to regulate this problem. This  limits the practical sizes
of the simulation.   The periodic boundary condition for the RVB
amplitudes  also presents a subtle problem.  Normally, we should use
those obtained by summing over a  discrete set of $\bk$'s in
accordance  with the finite size of the lattice.  If we did so,
there would be no sign problem for sizes up to $6\times 6$.  But
the energy would also get higher. We therefore use the amplitudes
calculated on a much larger lattice ($200\times 200$).
We then go to the smaller size and cut them  into a  {\em full}
hexgon (i.e. with inversion symmetry). Bonds outside are moved into
the hexgon  via the periodic boundary condition {\em except} for those that
are required to vanish by symmetry [e.g. for $i=(0,0)$ and $j=\pm
(n,-n)$  when the two axes are chosen to be (1,0) and ($-1,\sqrt{3}$)/2].
The  correlations as a function of distance can be obtained by
averaging over the six-fold  equivalences.

We wish to compare our  results simulated on finite lattices
with  those based on   exact diagonazations. On a $4\times 4$ lattice,
the  energy per bond is $E_{\mbox{RVB}}=-0.1754(3)$ vs.
$E_{\mbox{exact}}=-0.1782$ (cited from \cite{exact1}).  The difference
is  probably due to that the symmetry of the bonds are not very suited
on  this cluster. The result  on a $6\times 6$ lattice is presented in Table 1.
 One sees that they are in excellent agreement.
Perhaps, the only notable  difference between the two sets is at $n=3$.
The result also shows a nice approach to
(\ref{e33}),  i.e., a three-sublattice  magnetization.
{\em We stress that our wavefunction is parameter-free}.  These results
themselves  should be  of interest for variational approaches
{\em in  the presence of frustration}. Finally,
the small  difference in the  approximate  and  the true values of
$E_{\mbox{bond}}$ should justify  our analytic work.


Let us summarize the main results of this Letter.
Starting with a Schwinger-boson mean-field state that
includes both the antiferromagnetic and
ferromagnetic  fields, we have shown that a Gutzwiller projection leads
to optimizing the RVB state on the triangle lattice. This in turn
justifies the use of the two fields.   The ground-state energies of
the  optimized state were calculated rigorously  on finite lattices
and were  found to approach virtually the exact values. It is apparently much
better than all previous vaiational states\cite{KL,Huse}.
we have demonstrated, though our prediction for larger lattices is still based
on analytic approximation, that by properly optimizing the RVB pair
amplitudes one is most likely to get long-range correlations.
Namely, the system is predicted to possess a long-range order
within  {\em the RVB context}.


The author is indebted to  J. L. Lebowitz, Z. Y. Weng and T. K. Lee
for  helpful suggestions.  He also thanks the hospitality of
Rutgers University.   This work was supported by the National
Education Committee of China,  and by the US Air force Office of
Scientific Research through Grant No. 92-J-0115.


\newpage
\renewcommand{\thesection}{Table Caption}
\section{}

\begin{description}
\item[Table 1] The spin-spin correlation $C(n)$  as a function of distance
on the $6 \times 6$ lattice. The simulation takes up to $2\times
10^{9}$ runs and the statistical errors are about $\pm 0.002$. The
exact result is quoted  from Leung and Runge\cite{exact2}.
\end{description}

\newpage \renewcommand{\thepage}{Table 1}\Sp \vspace{1.5in}\large

\begin{center}\begin{tabular}{|c|c|c|c|c|c|c|}
\hline $n=$   &$1$ &$2$ &$3$ &$4$ &$5$ &$6$   \\ \hline
RVB &$-0.186$&$0.155$ &$-0.062$&$-0.065$ &$0.118$&$0.117$ \\ \hline
exact  &$-0.1868$ &$0.1535$&$-0.0548$ &$-0.0664$ &$0.1136$ &$0.1174$\\ \hline
\end{tabular}
\end{center}


\begin{thebibliography}{99}
\bibitem{class} G. H. Wannier, Phys. Rev. \u{79}, 357 (1950).
\bibitem{PWA} P. W. Anderson, Mater. Res. Bull. \u{8}, 153
	(1973); P. Fazekas and P. W. Anderson, Philos. Mag. \u{30},
	423 (1974).
\bibitem{KL} V. Kalmeyer and R. B. Laughlin, Phys. Rev. Lett. \u{59},
	2095 (1987).
\bibitem{Huse} D. A. Huse and V. Elser, Phys. Rev. Lett. \u{60},
	2531 (1988).
\bibitem{spin} Th. Jolicoeur and J. C. Le Guillou, Phys. Rev. B.
	\u{40} 2727 (1989); S. J. Miyake, J. Phys. Soc. Jpn. \u{61},
	983 (1992).
\bibitem{Large-N} S. Sachdev, Phys. Rev. B \u{45}, 12377 (1992).
\bibitem{SSB} P. Azaria, B. Delamotte and D. Mouhanna, Phys. Rev.
	Lett. \u{70}, 2483 (1993).
\bibitem{Ising} R. R. P. Singh, D. A. Huse, Phys. Rev. Lett. \u{68},
	1766 (1992).
\bibitem{high-T} N. Elstner, R. R. P. Singh, and A. P. Young, Phys.
	Rev. Lett. \u{71}, 1629 (1993).
\bibitem{exact0} B. Bernu, C. Lhuillier, and L. Pierre, Phys. Rev.
	Lett. \u{69}, 2590 (1992).
\bibitem{exact1} K. Yang, L. K. Warman, and S. M. Girvin, Phys. Rev.
	Lett. \u{70}, 2641 (1993).
\bibitem{exact2} P. W. Leung, K.  J. Runge, Phys. Rev. B \u{47}, 5861 (1993).
\bibitem{LDA}	S. Liang, B. Doucot, and P. W. Anderson, Phys. Rev. Lett.
	\u{61}, 365 (1988).
\bibitem{chen1}	Y.-C. Chen, Physica C \u{202}, 345 (1992);
	{\em ibid} \u{204}, 88 (1992).
\bibitem{chen2}	Y.-C. Chen, Phys. Lett. A \u{174}, 329 (1993);
	Y.-C. Chen and K. Xiu, {\em ibid} \u{181}, 373 (1993).
\bibitem{review} E. Manousakis, Rev. Mod. Phys. \u{63}, 1 (1991).
\bibitem{Super-C} G. Chen, H.-Q. Ding, and W. A. Goddard III, Phys. Rev.
	B  \u{46}, 2933 (1992).
\bibitem{Suth} B. Sutherland, Phys. Rev. B \u{37}, 3786 (1988); {\em ibid}
     \u{38}, 6855 (1988).
\bibitem{AA} The two decouplings were first introduced by D. P. Arovas
	and  A. Auerbach, Phys. Rev. B \u{38}, 316 (1988).
\end{thebibliography}
\end{document}